\documentclass[12pt]{article}
\usepackage{amsmath}
\usepackage{amssymb}
\usepackage{amsfonts}
\usepackage{amscd}
\usepackage[all]{xy}

\numberwithin{equation}{section}
\newcommand{\field}[1]{\mathbb{#1}}
\newcommand{\R}{\field{R}}

\title{
Two-step rational extensions of the harmonic oscillator: exceptional orthogonal polynomials and ladder operators}
\author{I Marquette$^1$, C Quesne$^{2}$\\ 
{\small $^1$ School of Mathematics and Physics, The University of Queensland,} \\ 
{\small Brisbane, QLD 4072, Australia}\\ 
{\small $^2$ Physique Nucl\'eaireTh\'eorique et Physique Math\'ematique,  Universit\'e Libre de Bruxelles,} \\ 
{\small Campus de la Plaine CP229, Boulevard~du Triomphe, B-1050 Brussels, Belgium}\\
{\small E-mail: i.marquette@uq.edu.au and cquesne@ulb.ac.be}}
\date{ }
\begin{document}
\baselineskip=22pt plus 1pt minus 1pt
%%%%%%%%%%%%%%%%%%%%%%%%%%%%%%%%%%%%%%%%%%%%%%%%%%%%%%%%%%
\maketitle

\begin{abstract} 
The type III Hermite $X_m$ exceptional orthogonal polynomial family is generalized to a double-indexed one $X_{m_1,m_2}$ (with $m_1$ even and $m_2$ odd such that $m_2 > m_1$) and the corresponding rational extensions of the harmonic oscillator are constructed by using second-order supersymmetric quantum mechanics. The new polynomials are proved to be expressible in terms of mixed products of Hermite and pseudo-Hermite ones, while some of the associated potentials are linked with rational solutions of the Painlev\'e IV equation. A novel set of ladder operators for the extended oscillators is also built and shown to satisfy a polynomial Heisenberg algebra of order $m_2-m_1+1$, which may alternatively be interpreted in terms of a special type of $(m_2-m_1+2)$th-order shape invariance property. 
\end{abstract}


\newpage
\section{Introduction}

During the last few years, a lot of research activity has been devoted to the construction of exceptional orthogonal polynomials (EOP), which are new complete and orthogonal polynomial systems extending the classical families of Hermite, Laguerre and Jacobi. In contrast with the latter, the former admit some gaps in the sequence of their degrees, the total number of them being referred to as the codimension.\par
%
%---------------------------------------------------------------------------------------------------------------------------
%
The first examples of EOP, the so-called Laguerre and Jacobi $X_1$ families, with codimension $m=1$, were proposed in the context of Sturm-Liouville theory \cite{gomez09, gomez10a}. Shortly thereafter, such EOP were applied to quantum mechanics and proved to be related to some exactly solvable (ES) rational extensions of well-known potentials \cite{cq08}. These new quantum potentials had the additional interesting property of enlarging the known list of translationally shape-invariant (SI) potentials, which was thought to be complete at that time \cite{genden, cooper}.\par
%
%--------------------------------------------------------------------------------------------------------------------------------------
%
In agreement with some previous works on algebraic deformations of SI potentials \cite{gomez04a, gomez04b}, it then appeared convenient to use a supersymmetric quantum mechanical (SUSYQM) method to construct some additional examples of quantum potentials related to $X_1$ EOP \cite{bagchi09} and to extend such an analysis by proposing some new Laguerre and Jacobi $X_2$ families \cite{cq09}. The transition from codimension $m=1$ to codimension $m=2$ was characterized by a growing complexity, since there appeared three distinct families of EOP (labelled as types I, II and III, respectively), the first two being associated with SI potentials and strictly isospectral to the partner conventional potential.\par
%
%-------------------------------------------------------------------------------------------------------------------------------
%
The next step was the obtention of type I and type II Laguerre and Jacobi $X_m$ EOP families and related potentials for arbitrarily large codimension $m$ \cite{odake09, odake10a} and the study of their properties \cite{odake10b, ho11a}. It then became clear that, in the $m=1$ special case, the polynomials of the type I and type II families happened to be proportional, hence leaving a single independent one. In addition, the $X_m$ EOP were shown to be obtainable through several equivalent approaches, such as the Darboux-Crum transformation \cite{sasaki10, gomez10b, gomez12a}, the Darboux-B\"acklund one \cite{grandati11a} and the prepotential method \cite{ho11b}.\par
%
%-----------------------------------------------------------------------------------------------------------------------
% 
Type III Laguerre $X_m$ EOP families were also studied in detail and shown to exist only for even codimension \cite{grandati11a}. Some results were also obtained for type III Jacobi ones \cite{ho11b}.\par
%
%------------------------------------------------------------------------------------------------------------------------
%
The development of these type III EOP families helped in understanding the absence of Hermite $X_1$ EOP families in the first works carried out on the subject \cite{gomez09, gomez10a}. Some Hermite-type polynomials had indeed been obtained in quantum mechanical models in the early 90s \cite{dubov92, dubov94} and re-discovered several times later on (see, e.g., \cite{gomez04a, junker97, carinena, fellows}), where they appeared in the vast domain of anharmonic oscillators constructed by SUSYQM techniques (see, e.g., \cite{sukumar, junker98, fernandez99, fernandez05, bermudez} and references quoted therein). It became clear \cite{grandati11b, ho11c} that such Hermite-type polynomials actually belonged to type III EOP families corresponding to even codimension and that no other Hermite EOP families could be constructed in a first-order SUSYQM approach (nor in any of the above-mentioned equivalent methods).\par
%
%--------------------------------------------------------------------------------------------------------------------------
%
Recently, it was shown that the list of $X_m$ EOP families (and associated potentials) was not exhaustive and that novel multi-indexed $X_{m_1,m_2,\ldots,m_k}$ families could be obtained through the use of multi-step Darboux algebraic transformations \cite{gomez12b}, the Crum-Adler mechanism \cite{odake11}, higher-order SUSYQM \cite{cq11a, cq11b} or multi-step Darboux-B\"acklund transformations \cite{grandati12a}. These works aimed at generalizing type I and type II $X_m$ EOP families and consequently at extending the class of SI potentials that are rational extensions of conventional ones.\par
%
%---------------------------------------------------------------------------------------------------------------------------
%
It is worth mentioning too that very recently some studies were devoted to the construction of new rational extensions of solvable potentials with a finite bound-state spectrum and to the study of the polynomials appearing in their wavefunctions \cite{grandati11b, ho11c, cq12a, cq12b, grandati12b}, as well as to some mathematical aspects of the theory \cite{gomez12c, sasaki12}.\par
%
%----------------------------------------------------------------------------------------------------------------------------
%
As far as the authors know, however, no attempt has been made so far at extending type III $X_m$ EOP families to multi-indexed ones. The purpose of the present paper is to start filling in the gap by considering the simplest example, namely Hermite $X_{m_1,m_2}$ EOP families that may occur in a second-order SUSYQM approach to the harmonic oscillator. Although the latter subject has a long history (see, e.g., \cite{fernandez99, fernandez05, bermudez} and references quoted therein), the EOP viewpoint will shed some new and interesting light on it.\par
%
%-----------------------------------------------------------------------------------------------------------------------------
%
Another purpose of this paper is to re-examine the construction of ladder operators for harmonic oscillator partners. We plan to show that for the potentials associated with Hermite $X_{m_1,m_2}$ EOP, there exists an alternative of the usual procedure that combines the oscillator creation and annihilation operators with the supercharges \cite{junker97, fernandez99, fernandez05} or combinations of supercharges \cite{bermudez, carballo, mateo}.\par
%
%---------------------------------------------------------------------------------------------------------------------------
%
In section 2, we review the construction of Hermite $X_m$ EOP families and associated harmonic oscillator rational extensions in first-order SUSYQM. We extend both of them to second-order SUSYQM in section 3. Ladder operators are then built for the corresponding potentials in section 4. Finally, section 5 contains the conclusion.\par
%
%========================================================================
%
\section{Harmonic oscillator rational extensions in first-order SUSYQM}

In first-order SUSYQM \cite{cooper}, one considers a pair of SUSYQM partners $H^{(\pm)}$,
\begin{equation}
\begin{split}
  & H^{(+)} = A^{\dagger} A = - \frac{d^2}{dx^2} + V^{(+)}(x) - E, \qquad H^{(-)} = A A^{\dagger} = 
      - \frac{d^2}{dx^2} + V^{(-)}(x) - E, \\
  & A^{\dagger} = - \frac{d}{dx} + W(x), \qquad A = \frac{d}{dx} + W(x), \\
  & V^{(\pm)}(x) = W^2(x) \mp W'(x) + E,  
\end{split}  \label{eq:SUSY}
\end{equation}
which intertwine with the first-order differential operators $A$ and $A^{\dagger}$ as $A H^{(+)} = H^{(-)} A$ and $A^{\dagger} H^{(-)} = H^{(+)} A^{\dagger}$. Here $W(x)$ is the superpotential, which can be expressed as $W(x) = - \bigl(\log \phi(x)\bigr)'$ in terms of a nodeless seed solution $\phi(x)$ of the initial Schr\"odinger equation
\begin{equation}
  \left(- \frac{d^2}{dx^2} + V^{(+)}(x)\right) \phi(x) = E \phi(x),  \label{eq:phi-E}
\end{equation}
$E$ is the factorization energy, assumed smaller than or equal to the ground-state energy of $V^{(+)}(x)$, and a prime denotes a derivative with respect to $x$.\par
%
%----------------------------------------------------------------------------------------------------------------------
%
Whenever both energies are equal and $\phi(x)$ is therefore the ground-state wavefunction of $V^{(+)}(x)$, the partner potential $V^{(-)}(x)$ has the same bound-state spectrum as $V^{(+)}(x)$, except for the ground-state energy which is removed (case {\it i}). For $E$ less than the ground-state energy, in which case $\phi(x)$ is a nonnormalizable eigenfunction of $V^{(+)}(x)$, the partner $V^{(-)}(x)$ has the same spectrum as $V^{(+)}(x)$ if $\phi^{-1}(x)$ is also nonnormalizable (case {\it ii}) or has an extra bound-state energy $E$ below the spectrum of $V^{(+)}(x)$ corresponding to the wavefunction $\phi^{-1}(x)$, if the latter is normalizable (case {\it iii}).\par
%
%-------------------------------------------------------------------------------------------------------------------------------
%
As well known, for the harmonic oscillator potential $V(x) = x^2$ ($x \in \R$), the corresponding Schr\"odinger equation has an infinite number of bound-state wavefunctions, which can be written as
\begin{equation}
  \psi_{\nu}(x) = {\cal N}_{\nu} H_{\nu}(x) e^{- \frac{1}{2} x^2}, \qquad {\cal N}_{\nu} = (\sqrt{\pi} 2^{\nu}
  \nu!)^{-1/2}, \qquad \nu=0, 1, 2, \ldots,  \label{eq:psi+}
\end{equation}
where $H_{\nu}(x)$ is a $\nu$th-degree Hermite polynomial \cite{abramowitz}. The associated bound-state energies are given by $2\nu+1$, $\nu=0$, 1, 2,~\ldots.\par
%
%-----------------------------------------------------------------------------------------------------------------------------
%
If one assumes
\begin{equation}
  V^{(+)}(x) = x^2  \label{eq:V+}
\end{equation}
and $E=1$ in (\ref{eq:SUSY}) and (\ref{eq:phi-E}), then $W(x) = x$ and the partner potential $V^{(-)}(x) = x^2 + 2$ is a translated oscillator, which reflects the SI of the harmonic oscillator \cite{cooper}.\par
%
%--------------------------------------------------------------------------------------------------------------------------------
% 
If now $E<1$, the only possible polynomial-type nodeless seed solutions of (\ref{eq:phi-E}) are provided by the functions \cite{fellows}
\begin{equation}
  \phi_m(x) = {\cal H}_m(x) e^{\frac{1}{2} x^2}, \qquad m=2, 4, 6, \ldots,  \label{eq:phi}
\end{equation}
where ${\cal H}_m(x)$ is a $m$th-degree pseudo-Hermite polynomial \cite{abramowitz}\footnote{This pseudo-Hermite polynomial may also be seen as a special case of generalized Hermite polynomial \cite{noumi, clarkson}.}, defined by
\begin{equation}
  {\cal H}_m(x) = (-{\rm i})^m H_m({\rm i}x) = m! \sum_{p=0}^{[m/2]} \frac{1}{p! (m-2p)!} (2x)^{m-2p}.
  \label{eq:pseudo}
\end{equation}
The corresponding factorization energies are
\begin{equation}
  E_m = -2m-1.  \label{eq:fact-energy}
\end{equation}
Since $\phi_m^{-1}(x)$ is normalizable on $\R$, we are in case {\it iii} of SUSYQM. It is worth observing here that for odd $m$, the function $\phi_m(x)$, as defined in (\ref{eq:phi}) and (\ref{eq:pseudo}), is also a solution of equation (\ref{eq:phi-E}) corresponding to (\ref{eq:fact-energy}), but as it vanishes at $x=0$, it does not qualify as a seed function in first-order SUSYQM.\par
%
%----------------------------------------------------------------------------------------------------------------------------------------------
%
The corresponding partner potential is now
\begin{equation}
  V^{(-)}(x) = x^2 - 2\left[\frac{{\cal H}''_m}{{\cal H}_m} - \left(\frac{{\cal H}'_m}{{\cal H}_m}\right)^2 + 1
  \right]  \label{eq:V-}
\end{equation}
and the spectra of $H^{(+)}$ and $H^{(-)}$ are
\begin{equation}
  E^{(+)}_{\nu} = 2(\nu+m+1), \qquad \nu=0, 1, 2, \ldots,
\end{equation}
and 
\begin{equation}
  E^{(-)}_{\nu} = 2(\nu+m+1), \qquad \nu=-m-1, 0, 1, 2, \ldots,
\end{equation}
respectively.\par
%
%-------------------------------------------------------------------------------------------------------------------
%
The excited-state wavefunctions of $H^{(-)}$ can be found by acting with the operator $A$, where
\begin{equation}
  W(x) = - x - \frac{{\cal H}'_m}{{\cal H}_m},
\end{equation}
on the wavefunctions $\psi^{(+)}_{\nu}(x) = \psi_{\nu}(x)$ of $H^{(+)}$. On using the Hermite and pseudo-Hermite polynomial identities given in appendix A, we obtain
\begin{equation}
\begin{split}
  & \psi^{(-)}_{\nu}(x) = {\cal N}^{(-)}_{\nu} \frac{e^{- \frac{1}{2} x^2}}{{\cal H}_m(x)} y^{(m)}_{\nu+m+1}(x),
       \qquad \nu=0, 1, 2, \ldots, \\
  & y^{(m)}_{\nu+m+1}(x) = - {\cal H}_m(x) H_{\nu+1}(x) - 2m {\cal H}_{m-1}(x) H_{\nu}(x), \\
  & {\cal N}^{(-)}_{\nu} = \frac{{\cal N}^{(+)}_{\nu}}{\sqrt{E^{(+)}_{\nu}}} = 
       \frac{{\cal N}_{\nu}}{\sqrt{2(\nu+m+1)}} = [\sqrt{\pi} 2^{\nu+1} (\nu+m+1) \nu!]^{-1/2}.   
\end{split}  \label{eq:psi-a}
\end{equation}
On the other hand, the ground-state wavefunction can be written as \cite{fellows}
\begin{equation}
  \psi^{(-)}_{-m-1}(x) = {\cal N}^{(-)}_{-m-1} \phi_m^{-1}(x), \qquad {\cal N}^{(-)}_{-m-1} = 
  \left(\frac{2^m m!}{\sqrt{\pi}}\right)^{1/2}.  \label{eq:psi-b}
\end{equation}
Hence the whole set of wavefunctions is given by
\begin{equation}
  \psi^{(-)}_{\nu}(x) = {\cal N}^{(-)}_{\nu} \frac{e^{- \frac{1}{2} x^2}}{{\cal H}_m(x)} y^{(m)}_n(x),
  \qquad n=\nu+m+1, \qquad \nu=-m-1, 0, 1, 2, \ldots,  \label{eq:psi-}
\end{equation}
where we define
\begin{equation}
  y^{(m)}_0(x) = 1.
\end{equation}
\par
%
%----------------------------------------------------------------------------------------------------------------------------------
%
The set of $n$th-degree polynomials $y^{(m)}_n(x)$, $n=0$, $m+1$, $m+2$,~\ldots, being orthogonal and complete with respect to the positive-definite measure $e^{-x^2} \bigl({\cal H}_m(x)\bigr)^{-2} dx$, is an EOP system $X_m$ of codimension $m$. As shown in appendix A, these polynomials satisfy the second-order differential equation
\begin{equation}
  \left[\frac{d^2}{dx^2} - 2\left(x + \frac{{\cal H}'_m}{{\cal H}_m}\right) \frac{d}{dx} + 2n\right] y^{(m)}_n(x) = 0.
  \label{eq:diffeqn1}
\end{equation}
\par
%
%=============================================================================
%
\section{Harmonic oscillator rational extensions in second-order SUSYQM}

In second-order SUSYQM (see, e.g., \cite{fernandez05, andrianov95, bagrov, samsonov, bagchi99, aoyama} and references quoted therein), the first-order differential operators $A^{\dagger}$, $A$ of equation (\ref{eq:SUSY}) are replaced by second-order ones
\begin{equation}
  {\cal A}^{\dagger} = \frac{d^2}{dx^2} - 2p(x) \frac{d}{dx} + q(x), \qquad {\cal A} = \frac{d^2}{dx^2} 
  + 2p(x) \frac{d}{dx} + 2p'(x) + q(x), 
\end{equation}
which intertwine with two partner Hamiltonians
\begin{equation}
  H^{(1)} = - \frac{d^2}{dx^2} + V^{(1)}(x), \qquad H^{(2)} = - \frac{d^2}{dx^2} + V^{(2)}(x), 
\end{equation}
as ${\cal A} H^{(1)} = H^{(2)} {\cal A}$ and ${\cal A}^{\dagger} H^{(2)} = H^{(1)} {\cal A}^{\dagger}$. As a consequence, the functions $p(x)$, $q(x)$ and the potentials $V^{(1,2)}(x)$ are constrained by the relations
\begin{equation}
\begin{split}
  & q(x) = - p' + p^2 - \frac{p''}{2p} + \left(\frac{p'}{2p}\right)^2 - \frac{c^2}{16p^2}, \\
  & V^{(1,2)}(x) = \mp 2p' + p^2 + \frac{p''}{2p} - \left(\frac{p'}{2p}\right)^2 + \frac{c^2}{16p^2},
\end{split}
\end{equation}
where $c$ is some integration constant.\par
%
%--------------------------------------------------------------------------------------------------------------------------------------
%
In the reducible case (corresponding to $c \in \R$) that we consider here, the operators ${\cal A}^{\dagger}$ and ${\cal A}$ can be factorized into products of first-order differential operators $A^{(i)\dagger} = - d/dx + W^{(i)}(x)$, $A^{(i)} = d/dx + W^{(i)}(x)$, $i=1$, 2, as ${\cal A}^{\dagger} = A^{(1)\dagger} A^{(2)\dagger}$ and ${\cal A} = A^{(2)} A^{(1)}$, while ${\cal A}^{\dagger} {\cal A}$ and ${\cal A} {\cal A}^{\dagger}$ become ${\cal A}^{\dagger} {\cal A} = \left(H^{(1)} + \frac{c}{2}\right) \left(H^{(1)} - \frac{c}{2}\right)$, ${\cal A} {\cal A}^{\dagger} = \left(H^{(2)} + \frac{c}{2}\right) \left(H^{(2)} - \frac{c}{2}\right)$, respectively. The first set of operators $A^{(1)\dagger}$, $A^{(1)}$ can be associated with partner Hamiltonians of the same type as those given in (\ref{eq:SUSY}), namely $H^{(+)} = A^{(1)\dagger} A^{(1)}$ and $H^{(-)} = A^{(1)} A^{(1)\dagger}$, with corresponding potentials $V^{(+)}(x)$, $V^{(-)}(x)$ and a factorization energy $E_1$. In the same way, for the second set of operators $A^{(2)\dagger}$, $A^{(2)}$, we may consider $\tilde{H}^{(+)} = A^{(2)\dagger} A^{(2)}$, $\tilde{H}^{(-)} = A^{(2)} A^{(2)\dagger}$, with respective potentials $\tilde{V}^{(+)}(x)$, $\tilde{V}^{(-)}(x)$ and a factorization energy $E_2$.\par
%
%-----------------------------------------------------------------------------------------------------------------------------
%
The relation between both approaches is given by $H^{(1)} = H^{(+)} + \frac{c}{2}$ and  $H^{(2)} = \tilde{H}^{(-)} - \frac{c}{2}$ with an intermediate Hamiltonian $H = - d^2/dx^2 + V(x) = H^{(-)} + \frac{c}{2} = \tilde{H}^{(+)} - \frac{c}{2}$, while the constant $c$ is related to the two factorization energies through $c = E_1 - E_2$ and the function $p(x)$ can be expressed in terms of the two superpotentials as $p(x) = (W^{(1)} + W^{(2)})/2$.\par
%
%--------------------------------------------------------------------------------------------------------------------------
%
These superpotentials can be obtained from seed eigenfunctions $\phi^{(1)}(x)$ and $\phi^{(2)}(x)$ of $H^{(+)}$ and $\tilde{H}^{(+)}$ as $W^{(i)}(x) = - \bigl(\phi^{(i)}(x)\bigr)'$, $i=1$, 2, respectively. This is equivalent to considering two seed eigenfunctions $\phi_1(x)$, $\phi_2(x)$ of the starting Hamiltonian $H^{(1)}$, such that $\phi^{(1)}(x) = \phi_1(x)$ and $\phi^{(2)}(x) = A^{(1)} \phi_2(x) = {\cal W}(\phi_1, \phi_2)/\phi_1$, where ${\cal W}(\phi_1, \phi_2)$ denotes the Wronskian of $\phi_1(x)$ and $\phi_2(x)$. In terms of the latter, one can write
\begin{equation}
  V^{(2)}(x) = V^{(1)}(x) - 2 \frac{d^2}{dx^2} \log {\cal W}(\phi_1, \phi_2).
\end{equation}
To be an acceptable quantum potential, $V^{(2)}(x)$ must be nonsingular in the domain of definition of $V^{(1)}(x)$, which implies that the Wronskian ${\cal W}(\phi_1, \phi_2)$ must be nodeless.\par
%
%----------------------------------------------------------------------------------------------------------------------------------
%
In the harmonic oscillator case, let us start with a pair of partner potentials $V^{(\pm)}(x)$ given by equations (\ref{eq:V+}) and (\ref{eq:V-}), where we replace $m$ by $m_1$ (and similarly in all equations of section 2). The first factorization energy $E_1 = - 2m_1 - 1$ being less than the ground-state one, a nodeless Wronskian ${\cal W}(\phi_1, \phi_2)$ can be obtained by assuming that $E_2 < E_1$ and that $\phi_2(x)$ has a single zero on $\R$ (see, e.g., \cite{fernandez05, bagrov, samsonov}). This can be achieved here by taking $E_2 = - 2m_2 - 1$ and $\phi_2(x) = \phi_{m_2}(x) = {\cal H}_{m_2}(x) e^{x^2/2}$ with $m_2$ odd  and such that $m_2 > m_1$. The Wronskian therefore becomes
\begin{equation}
  {\cal W}(\phi_1, \phi_2) = {\cal W}(\phi_{m_1}, \phi_{m_2}) = g_{\mu}(x) e^{x^2},
\end{equation}
where
\begin{equation}
  g_{\mu}(x) = {\cal W}({\cal H}_{m_1}, {\cal H}_{m_2}) = {\cal H}_{m_1} {\cal H}'_{m_2} - {\cal H}'_{m_1} 
  {\cal H}_{m_2} = 2 (m_2 {\cal H}_{m_1} {\cal H}_{m_2-1} - m_1 {\cal H}_{m_1-1} {\cal H}_{m_2})
  \label{eq:gmu}
\end{equation}
is a $\mu$th-degree polynomial with $\mu = m_1+m_2-1$, its highest-degree term being $2^{\mu+1} (m_2-m_1) x^{\mu}$.\par
%
%-------------------------------------------------------------------------------------------------------------------------
%
As a result, we can write the two superpotentials as
\begin{equation}
  W^{(1)}(x) = - x - \frac{{\cal H}'_{m_1}}{{\cal H}_{m_1}}, \qquad W^{(2)}(x) = - x 
  + \frac{{\cal H}'_{m_1}}{{\cal H}_{m_1}} - \frac{g'_{\mu}}{g_{\mu}},  \label{eq:W12}
\end{equation}
while the potentials obtained in the two equivalent approaches are given by
\begin{equation}
\begin{split}
  & V^{(+)}(x) = x^2, \\
  & V^{(-)}(x) = \tilde{V}^{(+)}(x) = x^2 - 2\left[\frac{{\cal H}''_{m_1}}{{\cal H}_{m_1}} 
        - \left(\frac{{\cal H}'_{m_1}}{{\cal H}_{m_1}}\right)^2\right] - 2, \\
  & \tilde{V}^{(-)}(x) = x^2 - 2 \left[\frac{g''_{\mu}}{g_{\mu}} - \left(\frac{g'_{\mu}}{g_{\mu}}\right)^2
        \right] - 4,  
\end{split}  \label{eq:SUSY-2}
\end{equation}
and
\begin{equation}
\begin{split}
  & V^{(1)}(x) = x^2 + m_1 + m_2 + 1, \\
  & V(x) = x^2 - 2\left[\frac{{\cal H}''_{m_1}}{{\cal H}_{m_1}} 
        - \left(\frac{{\cal H}'_{m_1}}{{\cal H}_{m_1}}\right)^2\right] + m_1 + m_2 - 1, \\
  & V^{(2)}(x) = x^2 - 2 \left[\frac{g''_{\mu}}{g_{\mu}} - \left(\frac{g'_{\mu}}{g_{\mu}}\right)^2
        \right] + m_1 + m_2 - 3,
\end{split}  \label{eq:SUSY-2bis}
\end{equation}
respectively.\par
%
%------------------------------------------------------------------------------------------------------------------------------------------
%
As examples of harmonic oscillator rational extensions obtained in second-order SUSYQM, let us quote
\begin{eqnarray}
  && V^{(2)}(x) = x^2 + \frac{32x^2}{4x^4 + 3} - \frac{384x^2}{(4x^4 + 3)^2} + 2 \quad \text{if $m_1=2$, 
      $m_2=3$},  \\
  && V^{(2)}(x) = x^2 + \frac{24(4x^4 + 5)}{8x^6 + 20x^4 + 10x^2 + 5} 
      - \frac{160(28x^4 + 20x^2 + 5)}{(8x^6 + 20x^4 + 10x^2 + 5)^2} + 4 \nonumber \\
  && \quad \text{if $m_1=2$, $m_2=5$}, \\
  && V^{(2)}(x) = x^2 + \frac{16(16x^6 + 28x^4 + 140x^2 - 749)}{16x^8 + 112x^6 + 168x^4 + 84x^2 + 21}
      \nonumber \\ 
  && \quad {}- \frac{896(1072x^6 + 1932x^4 + 1008x^2 + 273)}
     {(16x^8 + 112x^6 + 168x^4 + 84x^2 + 21)^2} + 6 \quad \text{if $m_1=2$, $m_2=7$}, \\
  && V^{(2)}(x) = x^2 + \frac{64(4x^6 + 4x^4 - 13x^2 + 112)}{16x^8 + 64x^6 + 120x^4 + 45}
      \nonumber \\ 
  && \quad {}- \frac{1024(328x^6 + 1020x^4 + 90x^2 + 315)}
     {(16x^8 + 64x^6 + 120x^4 + 45)^2} + 6 \quad \text{if $m_1=4$, $m_2=5$}, \\
  && V^{(2)}(x) = x^2 + \frac{8(80x^8 + 272x^6 + 352x^4 + 284x^2 + 239)}
      {32x^{10} + 272x^8 + 784x^6 + 840x^4 + 210x^2 + 105}
      \nonumber \\ 
  && \quad {}- \frac{64(13488x^8 + 68992x^6 + 103320x^4 + 40320x^2 + 4515)}
     {(32x^{10} + 272x^8 + 784x^6 + 840x^4 + 210x^2 + 105)^2} + 8 \nonumber \\
  && \quad \text{if $m_1=4$, $m_2=7$}.
\end{eqnarray}
It is worth observing here that some of these potentials have already made their occurrence in another context \cite{bermudez, mateo, marquette09} from rational solutions of the Painlev\'e IV equation \cite{gromak}. The first one of them, for instance, is a member of the shifted harmonic oscillator hyperconfluent hierarchy of \cite{mateo} and also appears in combination with a standard oscillator in the two-dimensional superintegrable system studied in example B of \cite{marquette09}.\par
%
%---------------------------------------------------------------------------------------------------------------------------
%
{}From SUSYQM, we directly get the energy spectra of $H^{(1)}$, $H$ and $H^{(2)}$ in the form
\begin{equation}
\begin{split}
  & E^{(1)}_{\nu} = 2\nu + m_1 + m_2 + 2, \qquad \nu=0, 1, 2, \ldots, \\
  & E_{\nu} = 2\nu + m_1 + m_2 + 2, \qquad \nu=-m_1-1, 0, 1, 2, \ldots, \\
  & E^{(2)}_{\nu} = 2\nu + m_1 + m_2 + 2, \qquad \nu=-m_2-1, -m_1-1, 0, 1, 2, \ldots.
\end{split}  \label{eq:energy}
\end{equation}
In particular, we note that, as expected, the two levels of energy $E^{(2)}_{-m_2-1} = m_1 - m_2$ and $E^{(2)}_{-m_1-1} = m_2 - m_1$, have been added below the usual harmonic oscillator spectrum starting with $E^{(2)}_0 = m_1 + m_2 + 2$.\par
%
%--------------------------------------------------------------------------------------------------------------------------------------
%
The corresponding wavefunctions $\psi^{(1)}_{\nu}(x)$ and $\psi_{\nu}(x)$ being given by the right-hand sides of equations (\ref{eq:psi+}) and (\ref{eq:psi-}) (with $m$ replaced by $m_1$), respectively, it only remains to determine those of $H^{(2)}$. Apart from the ground-state one, which is proportional to $\bigl(\phi^{(2)}(x)\bigr)^{-1}$, they can be most easily found by acting with the operator $A^{(2)}$ on $\psi_{\nu}(x)$ (see appendix A). The results read
\begin{equation}
  \psi^{(2)}_{\nu}(x) = {\cal N}^{(2)}_{\nu} \frac{e^{-\frac{1}{2}x^2}}{g_{\mu}(x)} y^{(\mu)}_n(x), \qquad
  n = \nu + \mu + 2, \qquad \nu = -m_2-1, -m_1-1, 0, 1, 2, \dots,  \label{eq:psi2a}
\end{equation}
where $y^{(\mu)}_n(x)$ is an $n$th-degree polynomial in $x$, defined by
\begin{equation}
\begin{split}
  & y^{(\mu)}_{m_1}(x) = {\cal H}_{m_1}, \\
  & y^{(\mu)}_{m_2}(x) = {\cal H}_{m_2}, \\
  & y^{(\mu)}_{m_1+m_2+\nu+1}(x) = (m_2-m_1) {\cal H}_{m_1} {\cal H}_{m_2} H_{\nu+1} 
        + 2[m_1(m_2+\nu+1) {\cal H}_{m_1-1} {\cal H}_{m_2} \\
  & \hphantom{y^{(\mu)}_{m_1+m_2+\nu+1}(x) =} {} - m_2(m_1+\nu+1) {\cal H}_{m_1} {\cal H}_{m_2-1}]
        H_{\nu}, \qquad \nu=0, 1, 2, \ldots,
\end{split} \label{eq:psi2b}
\end{equation}
and the normalization coefficient is given by
\begin{equation}
\begin{split}
  & {\cal N}^{(2)}_{-m_2-1} = \left(\frac{2^{m_2+1} m_2! (m_2-m_1)}{\sqrt{\pi}}\right)^{1/2}, \\
  & {\cal N}^{(2)}_{-m_1-1} = \left(\frac{2^{m_1-1} m_1!}{\sqrt{\pi} (m_2-m_1)}\right)^{1/2}, \\
  & {\cal N}^{(2)}_{\nu} = [\sqrt{\pi} 2^{\nu+2} (\nu+m_1+1) (\nu+m_2+1) \nu!]^{-1/2}, \qquad \nu=0, 1, 2, 
       \ldots.
\end{split}  \label{eq:psi2c}
\end{equation}
\par
%
%-----------------------------------------------------------------------------------------------------------------------------------
%
The $n$th-degree polynomials $y^{(\mu)}_n(x)$, $n=m_1$, $m_2$, $m_1+m_2+1$, $m_1+m_2+2$, \ldots, form an orthogonal and complete set with respect to the positive-definite measure $e^{-x^2} \bigl(g_{\mu}(x)\bigr)^{-2} dx$. We therefore conclude that such a set is a new EOP system $X_{m_1,m_2}$ of codimension $\mu = m_1+m_2-1$. The second-order differential equation satisfied by $y^{(\mu)}_n(x)$ can be written as (for details see appendix A)
\begin{equation}
  \left[\frac{d^2}{dx^2} - 2\left(x + \frac{g'_{\mu}}{g_{\mu}}\right) \frac{d}{dx} + 2n 
  + 2 \frac{\bar{g}_{\mu-2}}{g_{\mu}}\right] y^{(\mu)}_n(x) = 0,
  \label{eq:diffeqn2}
\end{equation}
where $\bar{g}_{\mu-2}(x)$ is a ($\mu-2$)th-degree polynomial defined by
\begin{equation}
  \bar{g}_{\mu-2}(x) = {\cal W}({\cal H}'_{m_1}, {\cal H}'_{m_2}).  \label{eq:gbar}
\end{equation}
\par
%
%------------------------------------------------------------------------------------------------------------------------------------------
%
As a last point in this section, it is interesting to note that the order of the seed functions $\phi_1$ and $\phi_2$ may be changed without affecting the final results, only intermediate potentials and Hamiltonians being modified. So, on taking $\bar{\phi}_1(x) = \phi_2(x) = {\cal H}_{m_2}(x) e^{x^2}$ ($m_2$ odd) and $\bar{\phi}_2(x) = \phi_1(x) = {\cal H}_{m_1}(x) e^{x^2}$ ($m_1$ even and such that $m_1 < m_2$), we obtain $\bar{A}^{(i)} = d/dx + \bar{W}^{(i)}(x)$, $\bar{W}^{(i)}(x) = - \bigl(\log \bar{\phi}^{(i)}(x)\bigr)'$, $i=1$, 2, with $\bar{\phi}^{(1)}(x) = \bar{\phi}_1(x) = \phi_2(x)$ and $\bar{\phi}^{(2)}(x) = \bar{A}^{(1)} \bar{\phi}_2(x) = {\cal W}(\phi_2, \phi_1)/\phi_2$. Equations (\ref{eq:SUSY-2}) and (\ref{eq:SUSY-2bis}) now become
\begin{equation}
\begin{split}
  & \bar{V}^{(+)}(x) = V^{(+)}(x) = x^2, \\ 
  & \bar{V}^{(-)}(x) = \tilde{\bar{V}}^{(+)}(x) = x^2 - 2\left[\frac{{\cal H}''_{m_2}}{{\cal H}_{m_2}} 
        - \left(\frac{{\cal H}'_{m_2}}{{\cal H}_{m_2}}\right)^2\right] - 2, \\
  & \tilde{\bar{V}}^{(-)}(x) = \tilde{V}^{(-)}(x) = x^2 - 2 \left[\frac{g''_{\mu}}{g_{\mu}} 
        - \left(\frac{g'_{\mu}}{g_{\mu}}\right)^2\right] - 4,
\end{split} 
\end{equation}
and
\begin{equation}
\begin{split}
  & \bar{V}^{(1)}(x) = V^{(1)}(x) = x^2 + m_1 + m_2 + 1, \\
  & \bar{V}(x) = x^2 - 2\left[\frac{{\cal H}''_{m_2}}{{\cal H}_{m_2}} 
        - \left(\frac{{\cal H}'_{m_2}}{{\cal H}_{m_2}}\right)^2\right] + m_1 + m_2 - 1, \\
  & \bar{V}^{(2)}(x) = V^{(2)}(x) = x^2 - 2 \left[\frac{g''_{\mu}}{g_{\mu}} 
        - \left(\frac{g'_{\mu}}{g_{\mu}}\right)^2\right] + m_1 + m_2 - 3,
\end{split}  
\end{equation}
respectively.\par
%
%------------------------------------------------------------------------------------------------------------------------------------------
%  
The resulting alternative factorizations ${\cal A} = A^{(2)} A^{(1)} = \bar{A}^{(2)} \bar{A}^{(1)}$ can be summarized in the following commutative diagram
\begin{equation}
\begin{CD}
  H^{(1)} @> \bar{A}^{(1)} >> \bar{H} \\
  @V A^{(1)} VV @VV \bar{A}^{(2)} V \\
  H @>> A^{(2)} > H^{(2)} 
\end{CD} \label{eq:commut}
\end{equation}
Although only formally defined (since $\bar{V}(x)$ is singular at $x=0$), the Hamiltonian $\bar{H}$ will prove very useful in the next section for constructing a new set of ladder operators for $H^{(2)}$.\par
%
%=======================================================================
%
\section{Ladder operators for harmonic oscillator rational extensions}

In the present section, we will start by reviewing the standard way of constructing ladder operators for $H^{(2)}$ in SUSYQM before introducing a new one and studying its properties.\par
%
%++++++++++++++++++++++++++++++++++++++++++++++++++++++++++++++++++++++
%
\subsection{\boldmath Standard ladder operators for $H^{(2)}$}

The usual procedure for building ladder operators for a SUSYQM partner Hamiltonian consists in combining those of the starting Hamiltonian with the supercharge operators \cite{fernandez99}. In the present case, from the standard harmonic oscillator raising and lowering operators
\begin{equation}
  a^{\dagger} = - \frac{d}{dx} + x, \qquad a = \frac{d}{dx} + x, \qquad [a, a^{\dagger}] = 2,  
\end{equation}
valid for $H^{(1)}$, those of $H^{(2)}$ are obtained in the form
\begin{equation}
  b^{\dagger} = {\cal A} a^{\dagger} {\cal A}^{\dagger}, \qquad b = {\cal A} a {\cal A}^{\dagger}
\end{equation}
and are therefore fifth-order differential operators.\par
%
%------------------------------------------------------------------------------------------------------------------------------
%
The operators $H^{(2)}$, $b^{\dagger}$ and $b$ satisfy the commutation relations
\begin{equation}
  [H^{(2)}, b^{\dagger}] = 2 b^{\dagger}, \qquad [H^{(2)}, b] = - 2 b, \qquad [b, b^{\dagger}] = P(H^{(2)}+2) 
  - P(H^{(2)}),  \label{eq:PHA1}
\end{equation}
where $P(H^{(2)})$ is a fifth-order polynomial in $H^{(2)}$, which can be factorized as
\begin{eqnarray}
  && P(H^{(2)}) = (H^{(2)}-m_1-m_2-2) (H^{(2)}-m_1+m_2-2) (H^{(2)}-m_1+m_2) 
        \nonumber \\
  && \quad {}\times (H^{(2)}+m_1-m_2-2) (H^{(2)}+m_1-m_2).  \label{eq:PHA2}
\end{eqnarray}
Hence they close a polynomial Heisenberg algebra (PHA) of fourth order.\par
%
%--------------------------------------------------------------------------------------------------------------------------------
%
Some information on the unitary irreducible representations of this PHA can be obtained by determining the zero modes of $b$ and $b^{\dagger}$ (i.e., the states that satisfy either $b\psi = 0$ or $b^{\dagger}\psi = 0$). Those of $b$ can be deduced from the vanishing of the norm of $b\psi$, which involves the average of the operator product $b^{\dagger}b = P(H^{(2)})$. Their energies are therefore given by $m_1-m_2$, $m_1-m_2+2$, $m_2-m_1$, $m_2-m_1+2$ and $m_1+m_2+2$. From (\ref{eq:energy}), we however know that a physical state can be associated with only three of these energies, namely $E^{(2)}_{-m_2-1} = m_1-m_2$, $E^{(2)}_{-m_1-1} = m_2-m_1$ and $E^{(2)}_0 = m_1+m_2+2$. In the same way, the zero modes of $b^{\dagger}$ can be obtained from the average of $bb^{\dagger} = P(H^{(2)}+2)$, leading to the energies $m_1-m_2-2$, $m_1-m_2$, $m_2-m_1-2$, $m_2-m_1$ and $m_1+m_2$. Here only $E^{(2)}_{-m_2-1} = m_1-m_2$ and $E^{(2)}_{-m_1-1} = m_2-m_1$ correspond to a physical state. We conclude that the PHA, defined in (\ref{eq:PHA1}) and (\ref{eq:PHA2}), has two one-dimensional unitary irreducible representations spanned by the singlets $\{\psi^{(2)}_{-m_2-1}\}$ and $\{\psi^{(2)}_{-m_1-1}\}$, respectively, and a single infinite-dimensional one spanned by $\{\psi^{(2)}_{\nu} \mid \nu=0, 1, 2, \ldots\}$.\par
%
%++++++++++++++++++++++++++++++++++++++++++++++++++++++++++++++++++++++++++++
%  
\subsection{\boldmath New ladder operators for $H^{(2)}$}

The construction of alternative ladder operators for $H^{(2)}$ is based upon the possibility of going from the first intermediate Hamiltonian $H$ of (\ref{eq:commut}) to the second (formal) one $\bar{H}$ (up to some additive constant) by a chain of $\ell = m_2-m_1$ first-order SUSYQM transformations characterized by the supercharges
\begin{equation}
\begin{split}
  & \hat{A}^{\dagger}_i = - \frac{d}{dx} + \hat{W}_i(x), \qquad \hat{A}_i = \frac{d}{dx} + \hat{W}_i(x), \\
  & \hat{W}_i(x) = x + \frac{{\cal H}'_{m_1+i-1}}{{\cal H}_{m_1+i-1}} 
       - \frac{{\cal H}'_{m_1+i}}{{\cal H}_{m_1+i}}, \qquad i=1, 2, \ldots, \ell.
\end{split}  \label{eq:Ai}  
\end{equation}
On defining 
\begin{equation}
  \hat{H}_i = - \frac{d^2}{dx^2} + x^2 - 2 \left[\frac{{\cal H}''_{m_1+i-1}}{{\cal H}_{m_1+i-1}}
  - \left(\frac{{\cal H}'_{m_1+i-1}}{{\cal H}_{m_1+i-1}}\right)^2\right] - 3, \qquad i=1, 2, \ldots, \ell+1,
\end{equation}
it is indeed straightforward to show that 
\begin{equation}
  \hat{A}^{\dagger}_i \hat{A}_i = \hat{H}_i, \qquad \hat{A}_i \hat{A}^{\dagger}_i = \hat{H}_{i+1} + 2, \qquad
  i=1, 2, \ldots, \ell,
\end{equation}
which implies that $\hat{H}_i$ and $\hat{H}_{i+1} + 2$ intertwine with $\hat{A}_i$ and $\hat{A}^{\dagger}_i$ as $\hat{A}_i \hat{H}_i = (\hat{H}_{i+1}+2) \hat{A}_i$ and $\hat{H}_i \hat{A}^{\dagger}_i = \hat{A}^{\dagger}_i (\hat{H}_{i+1}+2)$ for $i=1$, 2, \ldots, $\ell$. Since
\begin{equation}
  H = \hat{H}_1 + m_1 + m_2 + 2, \qquad \bar{H} = \hat{H}_{\ell+1} + m_1 + m_2 + 2,
\end{equation}
we infer that
\begin{equation}
  \hat{A}_{\ell} \cdots \hat{A}_2 \hat{A}_1 H = (\bar{H} + 2\ell) \hat{A}_{\ell} \cdots \hat{A}_2 \hat{A}_1,
  \qquad H \hat{A}^{\dagger}_1 \hat{A}^{\dagger}_2 \cdots \hat{A}^{\dagger}_{\ell} = \hat{A}^{\dagger}_1 
  \hat{A}^{\dagger}_2 \cdots \hat{A}^{\dagger}_{\ell} (\bar{H} + 2\ell),
\end{equation}
which proves the above assertion. This can be summarized in the following diagrammatic form
\begin{gather}
  \hat{H}_1 \xrightarrow{\hat{A}_1} \hat{H}_2+2 \xrightarrow{\hat{A}_2} \hat{H}_3+4 \xrightarrow{\hat{A}_3}
      \cdots \xrightarrow{\hat{A}_{\ell-1}} \hat{H}_{\ell}+2\ell-2 \xrightarrow{\hat{A}_{\ell}} \hat{H}_{\ell+1}+2\ell \\
  H \xrightarrow{\hat{A}_{\ell} \cdots \hat{A}_2 \hat{A}_1} \bar{H}+2\ell 
\end{gather}
\par
%
%----------------------------------------------------------------------------------------------------------------------------------------
%
This $\ell$th-order SUSYQM transformation from $H$ to $\bar{H}+2\ell$ can be combined with the two first-order ones going from $H^{(2)}$ to $H$ or from $\bar{H}$ to $H^{(2)}$ to provide some raising and lowering operators for $H^{(2)}$,
\begin{equation}
  c^{\dagger} = A^{(2)} \hat{A}_1^{\dagger} \hat{A}_2^{\dagger} \cdots \hat{A}_{\ell}^{\dagger}
  \bar{A}^{(2)\dagger}, \qquad c = \bar{A}^{(2)} \hat{A}_{\ell} \cdots \hat{A}_2 \hat{A}_1 A^{(2)\dagger},
  \label{eq:c}
\end{equation}
which are ($\ell+2$)th-order differential operators. From the set of intertwining relations established here as well as in section 3, it is indeed easy to prove that
\begin{equation}
  c H^{(2)} = (H^{(2)}+2\ell) c  \label{eq:ladder}
\end{equation}
or
\begin{equation}
  \xymatrixcolsep{5pc}\xymatrix@1{
  H^{(2)}  \ar@/_{10mm}/[rrr]^{c}  \ar[r]^{A^{(2)\dagger}} & H  \ar[r]^{\hat{A_{\ell}}\cdots\hat{A_{2}}\hat{A_ 
  {1}}}   & \bar{H}+2\ell \ar[r]^{\bar{A}^{(2)}}  & H^{(2)}+2\ell}
\end{equation}
Equation (\ref{eq:ladder}) may also be interpreted as a special type of ($\ell+2$)th-order SI, generalizing the third-order one considered elsewhere \cite{marquette09, andrianov00, ioffe}, to which it reduces in the case where $m_2 = m_1+1$ and therefore $\ell=1$.\par
%
%--------------------------------------------------------------------------------------------------------------------------------
%
The operators $H^{(2)}$, $c^{\dagger}$ and $c$ satisfy the commutation relations
\begin{equation}
  [H^{(2)}, c^{\dagger}] = 2\ell c^{\dagger}, \qquad [H^{(2)}, c] = - 2\ell c, \qquad [c, c^{\dagger}] = 
  Q(H^{(2)}+2\ell) - Q(H^{(2)}),  
\end{equation}
where 
\begin{equation}
  Q(H^{(2)}) = (H^{(2)}-3\ell) \Biggl[\prod_{i=1}^{\ell} (H^{(2)}-2m_1-\ell-2i)\Biggr] (H^{(2)}+\ell) 
\end{equation}
is a ($\ell+2$)th-order polynomial in $H^{(2)}$. Hence they close a PHA of ($\ell+1$)th order.\par
%
%--------------------------------------------------------------------------------------------------------------------------------
%
The zero modes of the annihilation operator $c$, deduced from the vanishing of the average of $c^{\dagger} c = Q(H^{(2)})$, correspond to the energies $-\ell$, $2m_1+\ell+2$, $2m_1+\ell+4$, \ldots, $2m_1+3\ell$ and $3\ell$ or, in other words, $m_1-m_2$, $m_1+m_2+2$, $m_1+m_2+4$, \ldots, $m_1+m_2+2\ell$ and $3m_2-3m_1$. Only the first $\ell+1$ of them, $E^{(2)}_{-m_2-1} = m_1-m_2$, $E^{(2)}_0 = m_1+m_2+2$, $E^{(2)}_1 = m_1+m_2+4$, \ldots, $E^{(2)}_{\ell-1} = m_1+m_2+2\ell$, can be associated with some physical state. On the other hand, the zero modes of the creation operator $c^{\dagger}$, obtained from the vanishing of the average of $c c^{\dagger} = Q(H^{(2)}+2\ell)$, correspond to the energies $-3\ell$, $2m_1-\ell+2$, $2m_1-\ell+4$, \ldots, $2m_1+\ell$ and $\ell$ or, equivalently, $3m_1-3m_2$, $3m_1-m_2+2$, $3m_1-m_2+4$, \ldots, $m_1+m_2$ and $m_2-m_1$, among which only the last one is a physical energy $E^{(2)}_{-m_1-1} = m_2-m_1$. We therefore obtain in this case one two-dimensional unitary irreducible representation spanned by the doublet of states $\{\psi^{(2)}_{-m_2-1}, \psi^{(2)}_{-m_1-1}\}$ and $\ell$ infinite-dimensional ones spanned by the states $\{\psi^{(2)}_{i+\ell j} \mid j=0,1,2,\ldots\}$ with $i=0$, 1, \ldots, $\ell-1$, respectively.\par
%
%------------------------------------------------------------------------------------------------------------------------------
%  
This is confirmed by determining the action of the annihilation operator $c$ on the physical wavefunctions $\psi^{(2)}_{\nu}(x)$, $\nu=-m_2-1$, $-m_1-1$, 0, 1, 2,~\ldots, given in equations (\ref{eq:psi2a})--(\ref{eq:psi2c}). This calculation, outlined in appendix B, leads to the results
\begin{equation}
\begin{split}
  & c \psi^{(2)}_{-m_2-1} = c \psi^{(2)}_0 = c \psi^{(2)}_1 = \cdots = c \psi^{(2)}_{m_2-m_1-1} = 0, \\
  & c \psi^{(2)}_{-m_1-1} = (m_2-m_1) \left(\frac{2^{m_2-m_1+2} m_2!}{m_1!}\right)^{1/2} 
        \psi^{(2)}_{-m_2-1}, \\
  & c \psi^{(2)}_{\nu} = \left(\frac{2^{m_2-m_1+2} \nu! (\nu+2m_1-m_2+1) (\nu+m_2+1)}{(\nu+m_1-m_2)!}
        \right)^{1/2} \psi^{(2)}_{\nu+m_1-m_2}, \\
  & \quad \nu=m_2-m_1, m_2-m_1+1, \ldots.  
\end{split} \label{eq:c-action}
\end{equation}
From (\ref{eq:c-action}), the action of the creation operator $c^{\dagger}$ can be directly obtained by Hermitian conjugation and in particular we have $c^{\dagger} \psi^{(2)}_{-m_1-1}=0$.\par
%
%======================================================================
%
\section{Conclusion}

In the present paper, we introduced a new EOP system by extending the type III Hermite $X_m$ EOP family \cite{fellows} to a double-indexed one $X_{m_1,m_2}$, where $m_1$ is even and $m_2$ is odd and greater than $m_1$. We showed that these new EOP can be easily expressed as linear combinations of mixed products of Hermite and pseudo-Hermite polynomials. Since their codimension $\mu = m_1+m_2-1$ is at least equal to four (corresponding to $m_1=2$ and $m_2=3$), such EOP do not appear in the recent classification of codimension two EOP systems obtainable from a classical system by a Darboux-Crum transformation \cite{gomez12c}. This illustrates the interest of exploring higher codimensions too.\par
%
%---------------------------------------------------------------------------------------------------------------------------------
%
We also constructed the related ES rational extensions of the harmonic oscillator and observed that some of them were already known and obtained from rational solutions of the Painlev\'e IV equation \cite{bermudez, mateo, marquette09, gromak}. This establishes an interesting link between such solutions and EOP.\par
%
%-----------------------------------------------------------------------------------------------------------------------
%
{}Furthermore, for these rational extensions of the harmonic oscillator, we proposed a new set of ladder operators giving rise to a PHA of ($m_2-m_1+1$)th order. In contrast with what happens for standard ladder operators \cite{fernandez99, fernandez05, bermudez, carballo, mateo}, the two states added below the harmonic oscillator spectrum belong to a single unitary irreducible representation of this PHA. Such an algebra also shows that the rationally-extended harmonic oscillator associated with the $X_{m_1,m_2}$ EOP family is endowed with a special type of ($m_2-m_1+2$)th-order SI, generalizing the third-order one considered in another context \cite{marquette09, andrianov00, ioffe}, to which it reduces whenever $m_2 = m_1+1$.\par
%
%-------------------------------------------------------------------------------------------------------------------------------
%
Some interesting problems for future study are the construction of type III multi-indexed  Hermite $X_{m_1,m_2,\ldots,m_k}$ EOP families and of the related harmonic oscillator extensions through the use of $k$th-order SUSYQM, as well as the generalization of the present approach to type III Laguerre and Jacobi EOP families. Two-dimensional superintegrable systems with higher-order integrals of motion, based on the one-dimensional systems presented here, would also be worth investigating along the lines of two recent works \cite{post, marquette12}, thereby generalizing special case B considered in \cite{marquette09}.\par
%
%=========================================================================
%
\section*{Acknowledgment}

The research of IM was supported by the Australian Research Council through Discovery Project DP110101414.\par
%
%===========================================================================
%
\section*{\boldmath Appendix A. Wavefunctions of SUSYQM partners}
\renewcommand{\theequation}{A.\arabic{equation}}
\setcounter{section}{0}
\setcounter{equation}{0}

The purpose of this appendix is to provide some details on the calculation of wavefunctions (\ref{eq:psi-a}), (\ref{eq:psi-b}) and (\ref{eq:psi2a})--(\ref{eq:psi2c}) of first- and second-order SUSYQM partners, respectively.\par
%
%----------------------------------------------------------------------------------------------------------------
%
In such a computation, we make a repeated use of the following Hermite and pseudo-Hermite identities
\begin{equation}
\begin{split}
  & H'_n = 2n H_{n-1}, \qquad H_{n+1} = 2x H_n - 2n H_{n-1}, \\
  & {\cal H}'_n = 2n {\cal H}_{n-1}, \qquad {\cal H}_{n+1} = 2x {\cal H}_n + 2n {\cal H}_{n-1},
\end{split}  \label{eq:A1}
\end{equation}
as well as of the differential equations
\begin{equation}
  H''_n - 2x H'_n + 2n H_n = 0, \qquad {\cal H}''_n + 2x {\cal H}'_n - 2n {\cal H}_n = 0.  \label{eq:A2}
\end{equation}
As a direct consequence of these relations, the Wronskian $g_{\mu}(x)$, defined in (\ref{eq:gmu}), fulfils the differential relations
\begin{equation}
  g'_{\mu} + 2x g_{\mu} = 2 (m_2-m_1) {\cal H}_{m_1} {\cal H}_{m_2}, \qquad g''_{\mu} + 2x g'_{\mu}
  - 2\mu g_{\mu} = 2 \bar{g}_{\mu-2},  \label{eq:A3}
\end{equation}
where $\bar{g}_{\mu-2}(x)$, defined in (\ref{eq:gbar}), can also be written as $\bar{g}_{\mu-2} = - 2m_1 {\cal H}_{m_1} {\cal H}'_{m_2} + 2m_2 {\cal H}'_{m_1} {\cal H}_{m_2}$.\par
%
%----------------------------------------------------------------------------------------------------------------------------
%
{}For the first-order SUSYQM partner, the excited-state wavefunctions (\ref{eq:psi-a}) can be derived from the SUSYQM property
\begin{eqnarray}
  \psi^{(-)}_{\nu}(x) & = &(E^{(+)}_{\nu} - E_m)^{-1/2} A \psi^{(+)}_{\nu}(x) = {\cal N}^{(-)}_{\nu} \left(
       \frac{d}{dx} - x - \frac{{\cal H}'_m}{{\cal H}_m}\right) H_{\nu}(x) e^{-\frac{1}{2} x^2} \nonumber \\
  & = & {\cal N}^{(-)}_{\nu} \frac{e^{-\frac{1}{2} x^2}}{{\cal H}_m(x)} \left[{\cal H}_m \left(\frac{d}{dx} - 2x
       \right) - {\cal H}'_m\right] H_{\nu}(x),
\end{eqnarray}
which directly leads to the desired result after using (\ref{eq:A1}). On the other hand, the normalized ground-state wavefunction (\ref{eq:psi-b}) has been taken from \cite{fellows}. It is worth noting here that in the same work \cite{fellows}, the EOP $y^{(m)}_{\nu+m+1}(x)$, $\nu=0$, 1, 2,~\ldots, given in (\ref{eq:psi-a}), have been instead written as a linear combination of $m+1$ Hermite polynomials. Finally, the second-order differential equation (\ref{eq:diffeqn1}), satisfied by $y^{(m)}_n(x)$, can be easily obtained by inserting equation (\ref{eq:psi-}) in the Schr\"odinger equation for $V^{(-)}(x)$ and using (\ref{eq:A2}).\par
%
%--------------------------------------------------------------------------------------------------------------------------------
% 
{}For the second-order SUSYQM partner, the ground-state wavefunction $\psi^{(2)}_{-m_2-1}(x)$ comes from the inverse of $\phi^{(2)}(x)$, leading to the expression of $y^{(\mu)}_{m_1}(x)$ in (\ref{eq:psi2b}). On the other hand, the excited-state wavefunctions result from the action of $A^{(2)}$ (with $W^{(2)}(x)$ given in (\ref{eq:W12})) on the intermediate Hamiltonian wavefunctions $\psi_{\nu}(x)$, $\nu=-m_1-1$, 0, 1, 2,~\ldots. On taking the first relation in (\ref{eq:A3}) into account, this yields
\begin{equation}
  A^{(2)} \psi_{\nu}(x) \propto 
  \frac{e^{-\frac{1}{2} x^2}}{g_{\mu}(x)}  
  \frac{1}{{\cal H}_{m_1}}\left[
  g_{\mu} \frac{d}{dx} - 2 (m_2-m_1) {\cal H}_{m_1} {\cal H}_{m_2}\right] y^{(m_1)}_{\nu+m_1+1}(x),
  \label{eq:A5}
\end{equation}
from which the expression of $y^{(\mu)}_{m_2}(x)$ (corresponding to $\nu=-m_1-1$) in (\ref{eq:psi2b}) follows. For higher values of $\nu$ ($\nu=0$, 1, 2,~\ldots), we have to insert definition (\ref{eq:gmu}) of $g_{\mu}(x)$ in the right-hand side of equation (\ref{eq:A5}). Among the resulting three terms inside square brackets, the first and third ones contain ${\cal H}_{m_1}$ as a factor, while the second one does not. After using the second relation in (\ref{eq:psi-a}) as well as equation (\ref{eq:A1}), this second term can however be rewritten as
\begin{equation}
  - 2m_1 {\cal H}_{m_1-1} {\cal H}_{m_2} \frac{d}{dx} y^{(m_1)}_{\nu+m_1+1} = 4m_1 (m_1+\nu+1)
  {\cal H}_{m_1} {\cal H}_{m_1-1} {\cal H}_{m_2} H_{\nu}.
\end{equation}
Hence all three terms in the numerator of (\ref{eq:A5}) contain the factor ${\cal H}_{m_1}$, which cancels the same in the denominator. On employing (\ref{eq:A1}) again, the remaining expression can be further transformed into the third relation in equation (\ref{eq:psi2b}).\par
%
%-----------------------------------------------------------------------------------------------------------------------------
%
The normalization coefficient ${\cal N}^{(2)}_{\nu}$, $\nu=-m_1-1$, 0, 1, 2,~\ldots, of the excited-state wavefunctions can be deduced from that of the intermediate Hamiltonian wavefunctions, from which such wavefunctions are derived, by multiplying the latter by the extra factor $\bigl(\tilde{E}^{(+)}_{\nu}\bigr)^{-1/2} = [2 (\nu+m_2+1)]^{-1/2}$. This yields the second and third relations in (\ref{eq:psi2c}), while the first one in the same is obtained from the second by observing that $\psi^{(2)}_{-m_2-1}$ and $\psi^{(2)}_{-m_1-1}$ have a similar form up to a permutation of $m_1$ and $m_2$.\par
%
%------------------------------------------------------------------------------------------------------------------------------
%
{}Finally, the differential equation (\ref{eq:diffeqn2}) comes from inserting equation (\ref{eq:psi2a}) in the Schr\"odinger equation for $V^{(2)}(x)$ and applying the second relation in (\ref{eq:A3}).\par
%
%=======================================================================
%
\section*{\boldmath \boldmath Appendix B. Action of the new annihilation operator $c$}
\renewcommand{\theequation}{B.\arabic{equation}}
\setcounter{section}{0}
\setcounter{equation}{0}

The purpose of this appendix is to prove that the action of the operator $c$, defined in (\ref{eq:c}), on the wavefunctions $\psi^{(2)}_{\nu}(x)$, $\nu=-m_2-1$, $-m_1-1$, 0, 1, 2,~\ldots, given in (\ref{eq:psi2a})--(\ref{eq:psi2c}), is provided by equation (\ref{eq:c-action}). Such a result actually agrees with SUSYQM predictions, but since the operators $\hat{A}_i$, $i=1$, 2, \ldots, $\ell$, and $\bar{A}^{(2)}$, used in definition (\ref{eq:c}), are only formally defined, we find it appropriate to make such an explicit check.\par
%
%---------------------------------------------------------------------------------------------------------------------
%
Since $A^{(2)\dagger}$ is a well-defined operator on $\R$, we may use its SUSYQM properties yielding
\begin{equation}
  A^{(2)\dagger} \psi^{(2)}_{\nu} = 
  \begin{cases}
    0 & \text{if $\nu=-m_2-1$}, \\
    \sqrt{2(\nu+m_2+1)} \psi_{\nu} & \text{if $\nu=-m_1-1$, 0, 1, 2, \ldots}.  
  \end{cases}
\end{equation}
This shows that $c \psi^{(2)}_{-m_2-1} = 0$ and
\begin{equation}
  c \psi^{(2)}_{\nu} \propto \bar{A}^{(2)} \hat{A}_{\ell} \cdots \hat{A}_2 \hat{A}_1 \frac{e^{-\frac{1}{2}x^2}}
  {{\cal H}_{m_1}} y^{(m_1)}_{\nu+m_1+1}, \qquad \nu=-m_1-1, 0, 1, 2, \ldots.  \label{eq:B2}
\end{equation}
\par
%
%--------------------------------------------------------------------------------------------------------------------------
%
To calculate the action of the product $\hat{A}_{\ell} \cdots \hat{A}_2 \hat{A}_1$ in (\ref{eq:B2}), we repeatedly employ the relation
\begin{eqnarray}
  \hat{A}_i \frac{e^{-\frac{1}{2}x^2}}{{\cal H}_{m_1+i-1}} y^{(m_1+i-1)}_{\nu+m_1+1} &=&
     \frac{e^{-\frac{1}{2}x^2}}{{\cal H}_{m_1+i-1}{\cal H}_{m_1+i}} \left[{\cal H}_{m_1+i} \frac{d}{dx}
     - 2 (m_1+i) {\cal H}_{m_1+i-1}\right] y^{(m_1+i-1)}_{\nu+m_1+1}  \nonumber \\
  & = & 2 (\nu-i+1) \frac{e^{-\frac{1}{2}x^2}}{{\cal H}_{m_1+i}} y^{(m_1+i)}_{\nu+m_1+1}, \qquad i=1, 2, \ldots,
     \ell,   
\end{eqnarray}
which results from (\ref{eq:psi-a}), (\ref{eq:Ai}), (\ref{eq:A1}) and the use of the intermediate equation
\begin{equation}
  \frac{d}{dx} y^{(m_1+i-1)}_{\nu+m_1+1} = - 2 (\nu+m_1+1) {\cal H}_{m_1+i-1} H_{\nu-i+1}, \qquad
  i=1, 2, \ldots, \ell.
\end{equation}
\par
%
%---------------------------------------------------------------------------------------------------------------------------
%
We finally arrive at
\begin{eqnarray}
  c \psi^{(2)}_{\nu} & \propto & \bar{A}^{(2)} \frac{e^{-\frac{1}{2}x^2}}{{\cal H}_{m_2}} 
      y^{(m_2)}_{\nu+m_1+1} \propto \left(\frac{d}{dx} - x + \frac{{\cal H}'_{m_2}}{{\cal H}_{m_2}} 
      - \frac{g'_{\mu}}{g_{\mu}}\right) \frac{e^{-\frac{1}{2}x^2}}{{\cal H}_{m_2}} y^{(m_2)}_{\nu+m_1+1},
      \nonumber \\
  && \quad \nu=-m_1-1, 0, 1, 2, \ldots,  
\end{eqnarray}
which can be easily shown to be given by
\begin{equation}
  c \psi^{(2)}_{\nu} \propto \frac{e^{-\frac{1}{2}x^2}}{g_{\mu}} y^{(\mu)}_{\nu-\ell+\mu+2}, \qquad 
  \nu=-m_1-1, 0, 1, 2, \ldots,
\end{equation}
on using (\ref{eq:psi-a}), (\ref{eq:A1}) and (\ref{eq:A3}).\par
%
%----------------------------------------------------------------------------------------------------------------------------
%
After collecting all coefficients, the result reads
\begin{equation}
  c \psi^{(2)}_{\nu} = \left\{ 
  \begin{array}{l}
    2\ell [(2m_1+2) (2m_1+4) \cdots (2m_1+2\ell)]^{1/2} \psi^{(2)}_{-m_2-1} 
         \qquad\text{if $\nu=-m_1-1$}, \\[0.2cm]
    0 \qquad\text{if $\nu=0, 1, \ldots, \ell-1$}, \\[0.2cm]
    [2\nu (2\nu-2) \cdots (2\nu-2\ell+2) (2\nu+2m_1-2\ell+2)]^{1/2}  \\[0.2cm]
    \quad {}\times (2\nu+2m_1+2\ell+2)^{1/2}\psi^{(2)}_{\nu-\ell} \qquad 
         \text{if $\nu=\ell, \ell+1, \ell+2, \ldots$},   
  \end{array}
  \right.
\end{equation}
which is equivalent to equation (\ref{eq:c-action}) when taking the definition of $\ell$ into account.\par
%
%============================================================================
%

\end{document}